\documentclass[prb,aps,twocolumn,showpacs,preprintnumbers,amsmath,amssymb,longbibliography]{revtex4-1}

\usepackage{graphicx}
\usepackage{dcolumn}

\usepackage[usenames,dvipsnames]{color}

\usepackage{soul}  

\newcommand\myuparrow{\mathord{\uparrow}}
\newcommand\mydownarrow{\mathord{\downarrow}}

\begin{document}

\def\EF{$E_\textrm{F}$}
\def\cred{\color{red}}
\def\cblue{\color{blue}}
\definecolor{dkgreen}{rgb}{0.31,0.49,0.16}
\def\cgreen{\color{dkgreen}}

\title{Nature of the Magnetic Interactions in Sr$_3$NiIrO$_6$}

\author{Turan Birol}
\affiliation{
Department of Physics and Astronomy, Rutgers University,
Piscataway, New Jersey 08854, USA
}
\affiliation{
Department of Chemical Engineering and Materials Science, University of Minnesota,
Minneapolis, Minnesota 55455, USA
}
\author{Kristjan Haule}
\affiliation{
Department of Physics and Astronomy, Rutgers University,
Piscataway, New Jersey 08854, USA
}
\author{David Vanderbilt}
\affiliation{
Department of Physics and Astronomy, Rutgers University,
Piscataway, New Jersey 08854, USA
}

\date{\today}

\begin{abstract}
Iridates abound with interesting magnetic behaviours because of their strong spin-orbit 
coupling. Sr$_3$NiIrO$_6$ brings together the spin-orbital entanglement of the 
Ir$^{4+}$ ion with a 3$d$ Ni cation and a one-dimensional crystal structure. 
It has a ferrimagnetic ground state with a 55\,T coercive field. 
We perform a theoretical study of the magnetic interactions in 
this compound, and elucidate the role of anisotropic symmetric exchange as the source of 
its strong magnetic anisotropy. 
Our first-principles calculations reproduce the magnon spectra of this compound and predict a 
signature in the cross sections that can differentiate the anisotropic 
exchange from single-ion anisotopy. 
\end{abstract}


\maketitle
\section{Introduction}
Oxides of 5$d$ transition metals, especially iridates, are at the center of recent interest because 
of their strong spin-orbit coupling (SOC).\cite{witczak-krempa2014,rau2016} SOC gives rise to novel 
phenomena such as anisotropic pseudodipolar magnetic exchange 
interactions.\cite{khaliullin2005} These interactions in turn lead to phases such as the 
quantum spin liquid in the Kitaev model, which might be realized in
honeycomb iridates. \cite{hermanns2017}

The magnetic behaviour of complex oxides with multiple inequivalent transition-metal cations can also be 
very rich, especially when the transition metals come from different rows of the periodic table. 
For example, the 3$d$-5$d$ double perovskites are known to display incommensurate antiferromagnetism, multiferroicity, 
magnetoresistance, half-metallic ferrimagnetism, independent ordering of interpenetrating magnetic 
lattices, and very often high ordering temperatures.\cite{lezaic2011, rolfs2017, erten2011, morrow2013, das2011}
Novel phenomena are still being discovered in these systems, such as the recent demonstration
of magnetic interactions in Ca$_2$CoOsO$_6$ and Ca$_2$NiOsO$_6$ that break
the Goodenough-Kanamori-Anderson (GKA) rules.\cite{morrow2016}

Another structural family of compounds that contain two different transition-metal cations is 
the A$_3$MM'O$_6$ chain compounds with the K$_4$CdCl$_6$
crystal structure.\cite{K4CdCl6_Structure} Many members of this family exhibit 
phenomena such as multiferroicity, unexpectedly strong magnetic anisotropy, colossal magnon gaps, 
superparamagnetic-like behaviour, and partially disordered 
antiferromagnetism.\cite{wu2009, yin2013, sampathkumaran2002, niitaka2001, mikhailova2014, nguyen1994}
Ferrimagnetic Sr$_3$NiIrO$_6$,\cite{nguyen1995} a member of this family, displays Ising-like magnetic anisotropy, a record-breaking magnetic 
coercive field,\cite{singleton2016} and a colossal spin wave gap.\cite{lefrancois2016, toth2016} 
All of these are surprising observations because neither Ni$^{2+}$ nor 
Ir$^{4+}$ should have strong single-ion magnetic anisotropy (SIA).  

In this work, we approach the magnetic interactions in Sr$_3$NiIrO$_6$ from first principles and elucidate the microscopic 
mechanism behind its magnetism. 
Our main result is that the effective magnetic interaction between nearest neighbor ions' moments ${\bf M}_i$ is 
symmetric but \textit{radically anisotropic}. In other words, while the energy expression does not contain 
anti-symmetric cross product terms, it has opposite 
signs for different components of the magnetic moments. 
This radically anisotropic symmetric interaction can be written as 
\begin{equation}
E= J_{\parallel} M_{i,z} M_{i+1,z} + J_{\perp} \left(M_{i,x} M_{i+1,x} + M_{i,y} M_{i+1,y}\right)
\end{equation}
where $J_{\parallel}$ and $J_{\perp}$ have opposite signs.
The strong Ising-like behaviour observed in this compound can be explained by this energy expression without
a single ion anisotropy (SIA) term.
%
We reproduce all three important qualitative aspects of the experimental magnon spectra (a small bandwidth, a 
much larger splitting of the bands, and a gap comparable to the splitting of magnon branches),\cite{lefrancois2016, toth2016}
using this model with parameters fitted to first-principles calculations without any fine tuning of parameters. 
%
We predict that a corollary of the radically anisotropic exchange is the flipping of the oscillation patterns of optical and acoustic magnons at the 
$\Gamma$ point, and 
conclude by putting forward a signature in the magnon-creation neutron-scattering cross sections 
that can be used to experimentally differentiate between the anisotropic exchange scenario and the commonly used isotropic exchange with 
strong SIA. 

Sr$_3$NiIrO$_6$ has been previously studied theoretically by multiple authors, starting with Vajenine and Hoffmann's Hueckel 
calculations.\cite{vajenine1996}
Zhang et al., using Density Functional Theory (DFT), underlined the importance of the SOC, and 
reported considerable orbital moments for both transition metals.\cite{zhang2010} Sarkar et al.\ 
verified the presence of large orbital moments.\cite{sarkar2010}
Ou and Wu pointed out the importance of SOC in altering the orbital configuration of Ir, and 
found that it is responsible for the intrachain ferrimagnetic order.\cite{ou2014}
Most recently, Gordon et al.\ elucidated a connection between the magnetic exchange interactions 
and the Ising behaviour in Sr$_3$NiIrO$_6$.\cite{gordon2016} 
Our work goes beyond these first-principles calculations, and  in addition to explaining the microscopic 
mechanism of the anisotropic exchange interaction in this 3$d$-5$d$ system, bridges the gap between the first-principles 
calculations and experimental observations by calculating the 
magnetic interaction parameters from first principles and reproducing the experimentally observed magnon spectra.

\section{Magnetic Structure}

\begin{figure}
\includegraphics[width=.9\linewidth]{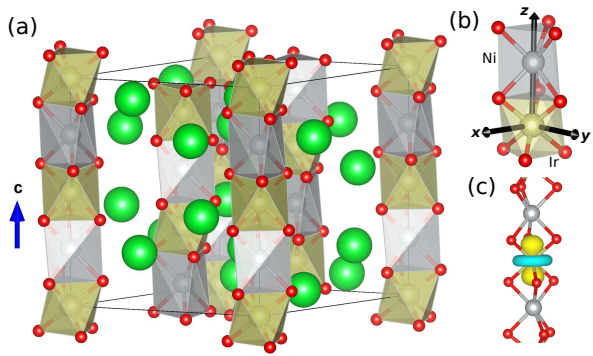}
\caption{(Color Online)
(a) Crystal structure of Sr$_3$NiIrO$_6$ 
(b) The local axes used for the Ir ions. (c) The $|3z^2-r^2\rangle$ orbital on an Ir ion. }
\label{fig:struct}
\end{figure}

The structure of Sr$_3$NiIrO$_6$ consists of parallel one-dimensional chains of 
alternating face-sharing NiO$_6$ and IrO$_6$ polyhedra as shown in Fig.~\ref{fig:struct}(a).\cite{K4CdCl6_Structure, nguyen1995} 
There are two different energy scales for magnetic couplings along the 
$c$ axis (intrachain) and in the $ab$ plane (interchain). Similarly, there are 
two temperature scales for magnetism. At $T_2$, intrachain magnetic order sets in. The temperature scale for the interchain magnetic 
order, $T_1$, is usually about an order of magnitude smaller than $T_2$. This is because there are no good superexchange 
paths that connect magnetic moments in different chains, but also because the chains form a frustrated triangular lattice. In Sr$_3$NiIrO$_6$, $T_2=75$ K, and the intrachain order is ferrimagnetic: 
both the Ni and Ir moments are aligned along the $c$ axis (chain direction), but are anti-parallel to each 
other.\cite{lefrancois2014} The interchain order below $T_1=17$ K is still under debate: 
neutron data is consistent with both the so-called partially disordered and the amplitude-modulated antiferromagnetic 
arrangements of the ferrimagnetic chains.\cite{lefrancois2014} In this work, we focus only on the intrachain interactions, 
and do not address the question of interchain magnetic order.

\begin{figure}
\includegraphics[width=1.0\linewidth]{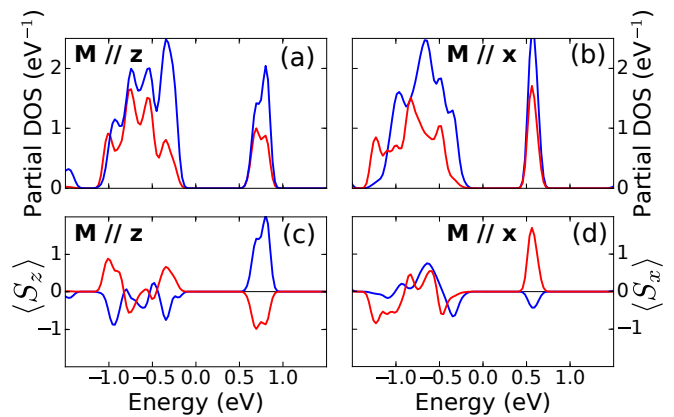}
\caption{(Color Online)
(a) Densities of states of the Ir ion projected onto the $d$ orbitals in the FiM state with magnetic moments along the $z$ axis 
(magnetic ground state). 
Red curve is the DOS projected onto the $|3z^2-r^2\rangle$ orbital, and the blue curve is the sum of the projected DOS's onto the other four d orbitals.
(b) Same quantity as in (a), but in the FM state with magnetic moments along the $x$ axis. (c) 
Energy resolved expectation value of the $z$ component of spin, $\langle S_z \rangle$ for the $d$ orbitals of the Ir ion  in the 
FiM state with magnetic moments along the $z$ axis. 
(d) Same quantity as in (c), but in the FM state with magnetic moments along the $x$ axis. 
}\label{fig:dosIr}
\end{figure}

In Fig.~\ref{fig:dosIr}(a) we plot the energy-resolved density of states on the Ir ion, with the 
projections on the $|3z^2-r^2\rangle$ orbital plotted separately, from DFT+$U$+SOC calculations. 
In our choice of coordinate axes, shown in Fig.~\ref{fig:struct}(b), 
the $|3z^2-r^2\rangle$ orbital of Ir has $t_{2g}$-like character with lobes extended towards the nearest neighbor Ni 
ions as shown in Fig.~\ref{fig:struct}(c).
This makes it the most important orbital for the superexchange interactions.\cite{Note1}
The Ir$^{4+}$ cation has five $d$ electrons in its valence shell. Its unoccupied $e_g$ 
states lie between 3 and 4\,eV (not shown), and it has a single Ir $t_{2g}$ hole between 0.5 and 1.0\,eV. This hole 
has 36\% $|3z^2-r^2\rangle$ character, and a nontrivial spin characteristic: its $|3z^2-r^2\rangle$ contribution has the 
opposite spin relative to the rest of the hole, as seen in the energy-resolved spin expectation value $\langle S_z\rangle$ 
in Fig.~\ref{fig:dosIr}(c). 
This can be explained by the strong SOC of the Ir ion: the hole on the Ir does not have a definite spin, but it has $J_\textrm{eff}=1/2$ 
character and can be thought to have a corresponding \textit{pseudo}spin. 
The $J_\textrm{eff}=1/2$ orbitals with pseudospin in the $\mp \hat{z}$ directions are 
\begin{equation}
|J_{1/2},\uparrow\rangle = \frac{1}{\sqrt{|\gamma|^2+2}}  \left(i\gamma|A,\downarrow\rangle + 
\sqrt{2} |E^{+},\uparrow\rangle\right)\label{equ:Jefftri1}
\end{equation}
and
\begin{equation}
|J_{1/2},\downarrow\rangle = \frac{1}{\sqrt{|\gamma|^2+2}}  \left(i\gamma|A,\uparrow\rangle + 
\sqrt{2} |E^{-},\downarrow\rangle\right)\label{equ:Jefftri2}
\end{equation}
where $|A\rangle=|3z^2-r^2\rangle$ and $|E^\mp\rangle$ are the $t_{2g}$-like orbitals that are split by the trigonal field. 
(In the absence of the trigonal crystal field, $\gamma=1$.) As a result, \textit{the spin moment on the 
$|A\rangle = |3z^2-r^2\rangle$ orbital is opposite to the total spin moment (as well as the pseudospin moment) of the Ir ion when it is along the $z$ direction.}

\section{Magnetic interactions}
The effective magnetic Hamiltonian for the Ir ion with an electron in the $J_\textrm{eff}=1/2$ orbital needs to
be built using not the spin, but rather the pseudospin of the electron.
The SOC reduces the orbital degeneracy in iridates, but the magnetic Hamiltonians are usually more complicated 
and may involve anisotropic exchange interactions which couple different components of pseudospins with different 
strengths.\cite{jackeli2009,khaliullin2005}
Anisotropic exchange interactions can set a preferred axis for the pseudospin moments
and give rise to other effects that are usually ascribed to SIA. 
For example, the magnon gap observed in Sr$_3$Ir$_2$O$_7$ is explained by the exchange 
anisotropy between the Ir$^{4+}$ ions.\cite{kim_magnon2012}

The GKA rules\cite{goodenough1963} for the signs of the exchange interactions do not 
directly apply to the pseudospins since the orbital degree of freedom is entangled with spin.\cite{morrow2016} 
Instead, we need to consider the individual orbital components of the $J_\textrm{eff}=1/2$ spin-orbitals and the 
interactions between them.  
A tight binding model constructed using the ab-initio Wannier functions\cite{WANNIER_REVIEW, marzari1997} that only includes the 
Ni $d$ and Ir $t_{2g}$ orbitals shows that the largest 
hopping is between the $|A\rangle$ orbital on the Ir and the similar $|3z^2-r^2\rangle$ on the Ni as expected in this 
face-sharing polyhedral geometry.\footnote{See the supplemental material, which includes Refs. 
\onlinecite{VASP1, VASP2, PAW1, PAW2, PBEsol, DFTU:Dudarev, hoffmann1976, bilbao1, fazekas1999, 
lovesey1986, liu2015, solovyev1996},  for details.}

The DOS projected onto Ni shows\cite{Note1} that the $|3z^2-r^2\rangle$ orbital on Ni$^{2+}$ 
is fully occupied. The superexchange process in which a Ni electron is excited to an Ir 
$|A\rangle$ orbital is possible only if the electron has opposite spin to the spin
on the Ir $|A\rangle$ orbital, and 
provides an energy gain proportional to the Ni on-site Hund's coupling if the Ni spin is parallel to that on the Ir $|A\rangle$ 
orbital.\cite{goodenough1963}
This implies that there is a ferromagnetic coupling between the Ni spin
and the spin on the Ir $|A\rangle$ orbital. 
Since the total spin expectation value $\langle S_z\rangle$ of the Ir ion is opposite to the spin on the Ir 
$|A\rangle$ orbital, this superexchange provides an antiferromagnetic (ferrimagnetic) coupling between the total magnetic moments 
of the Ni and the Ir ions. The inclusion of other orbitals in this argument\cite{Note1} does not change
the sign of this coupling, which explains the ferrimagnetic ground state observed in Sr$_3$NiIrO$_6$. 

In our picture, then, this antiferromagnetic interaction, which emerges from a ferromagnetic superexchange, is a 
direct result of the strong SOC on the Ir ion. In order to test this claim further,
we repeated a similar 
DFT calculation with SOC turned off, and could stabilize only the FM configuration. This is consistent 
with Ref.~[\onlinecite{sarkar2010}] where SOC was not taken into account and consequently only the FM 
order was stabilized. 
A similar point about different orbitals contributing to superexchange in a nontrivial way due to SOC 
is also made in Ref. \onlinecite{yin2013}, where a ferromagnetic exchange anisotropy stemming from an 
antiferromagnetic exchange interaction in Sr$_3$CuIrO$_6$ is studied.

%
The Ir $|A\rangle$ orbital has opposite spin to that of the $|E^\mp\rangle$ orbitals only when the Ir pseudospin 
is along $\hat{z}$. If the Ir magnetic moment is in another direction, this condition will no longer be satisfied. 
For example, the $J_\textrm{eff}=1/2$ state with pseudospin parallel to 
$\hat{x}$, 
\begin{equation}
|J_{1/2},\uparrow_x\rangle= \left(|J_{1/2},\uparrow\rangle +  |J_{1/2},\downarrow\rangle\right)/\sqrt{2}
\end{equation}
has the form
\begin{align}
|J_{1/2},\uparrow_x\rangle &= \Big( i\gamma\sqrt{2}|A,\uparrow_x\rangle +   |E^+,\uparrow_x\rangle + |E^-,\uparrow_x\rangle \nonumber\\
  & \hspace{.5cm} +|E^+,\downarrow_x\rangle  -|E^-,\downarrow_x\rangle  \Big)/\sqrt{2(|\gamma|^2+2)}
\end{align}
where $|\myuparrow_x\rangle=(|\myuparrow\rangle + |\mydownarrow \rangle)/\sqrt{2}$
and $|\mydownarrow_x\rangle=(|\myuparrow\rangle - |\mydownarrow \rangle)/\sqrt{2}$.
The spin on the $|A\rangle$ orbital is now parallel to the pseudospin of the $|J_\textrm{eff},\uparrow_x\rangle$ 
orbital. DFT results with the spins aligned in the $x$-$y$ plane, shown in Fig.~\ref{fig:dosIr}(c) and (d), 
are consistent with this observation:
the $\langle S_x \rangle$ of the hole on $|A\rangle$ is parallel to the minority spin direction. In this case, the ferromagnetic 
superexchange between the Ni ion and the Ir $|A\rangle$ orbital should give rise to a ferromagnetic coupling between 
the magnetic moments of these ions. In other words, the effective interaction between the
magnetic moments $M$ on the nearest-neighbor Ni-Ir atoms
is anisotropic and has the form $E= J_\parallel M_z^\textrm{Ir} M_z^\textrm{Ni}+J_\perp M_{xy}^\textrm{Ir} M_{xy}^\textrm{Ni}$
with $J_{\parallel} > 0$ but $J_\perp<0$. 

\begin{figure}
\includegraphics[width=.9\linewidth]{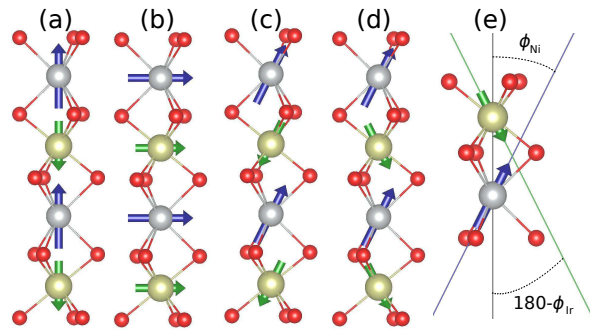}
\caption{(Color Online)
(a) Ferrimagnetic state with moments along the
${z}$ axis, which is the lowest-energy state
(FiM-$z$: $\phi_{\rm Ni-Ir}=180^\circ$, $\phi_{\rm Ni}=0^\circ$,
$\phi_{\rm Ir}=180^\circ$).
(b) Ferromagnetic state with moments along the ${x}$ axis
(FM-$x$: $\phi_{\rm Ni-Ir}=0^\circ$, $\phi_{\rm Ni}=90^\circ$, $\phi_{\rm Ir}=90^\circ$).
(c) Possible ferrimagnetic intermediate state ($\phi_{\rm Ni-Ir}=180^\circ$).
(d) Observed, canted ferrimagnetic intermediate state ($\phi_{\rm Ni-Ir}<180^\circ$).
(e) Definitions of $\phi_{\rm Ni}$ and $\phi_{\rm Ir}$.
}
\label{fig:magneticstates}
\end{figure}

DFT calculations provide estimates of $J_\perp$ and $J_\parallel$ that support this claim.
We adopt the standard approach of initiating
the DFT calculations in different magnetic configurations 
to estimate the energy differences between various 
magnetic states.
However, especially in noncollinear calculations, it is not always possible to stabilize the system in
the desired local energy minima if the system is very far from its groundstate 
\footnote{Some other studies (Refs.~\onlinecite{zhang2010, ou2014}) found similar behavior.}.
When we initiate our DFT calculation with spins parallel to $\hat{z}$, all of our calculations (even those initiated with FM order) 
converge to the ferrimagnetic configuration (\mbox{FiM-$z$}, Fig.~\ref{fig:magneticstates}(a)).
On the other hand, we could not stabilize a FiM state with spins in 
the $x$-$y$ plane; the only state we could stabilize with moments in the $x$-$y$ plane
is the ferromagnetic one (\mbox{FM-$x$}, Fig.~\ref{fig:magneticstates}(b)).

To gain information about the magnetic interactions, then, we compute the energy at both of the 
energy minima (FM-$x$ and FiM-$z$, shown in Fig. \ref{fig:magneticstates}(a) and (b)), where it is possible 
to converge the electronic state to a very high precision, 
and also at several intermediate states where the magnetic moments are tilted. 
We do not observe any local minima in the vicinity of these intermediate states, but the slope of the electronic energy surface is so small 
(changing by less than about $10^{-5}$\,eV/atom from one self-consistent iteration to the next)
that we believe it is well justified to estimate the energy of these
intermediate states in this way.\footnote{Keeping the calculation running 
until selfconsistency results 
in these intermediate states eventually converging to either FM-$x$
or FiM-$z$.}

\begin{figure}
\includegraphics[width=0.7\linewidth]{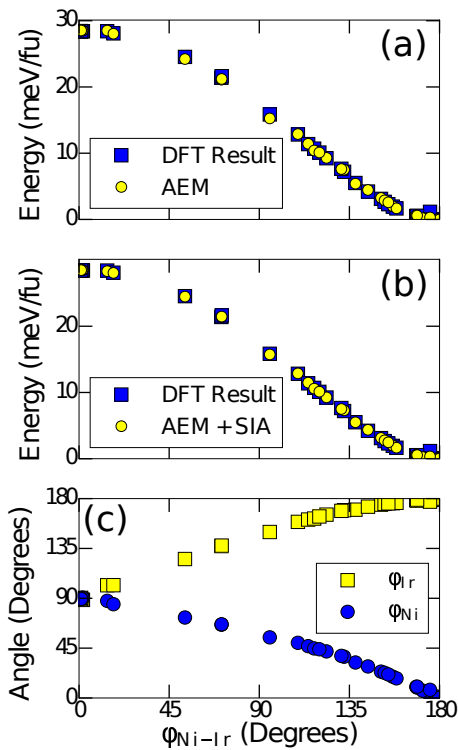}
\caption{(Color Online)
First-principles results for the intrachain magnetic interaction between Ni and Ir atoms. 
(a) Energy as a function of the angle between the magnetic moments of nearest-neighbor atoms. 
Blue squares are the energy calculated from DFT, and yellow circles are the energy obtained 
from the fitted anisotropic exchange model (AEM). 
(b) Same as in (a), but yellow circles are the energy obtained from the fitted model with 
anisotropic exchange and SIA on the Ni ion. 
(c) The angle that the Ir and Ni magnetic moments make with the [001] axis as a function 
of the angle between the magnetic moments of nearest-neighbor atoms.
}
\label{fig:magnetics}
\end{figure}

We summarize our results in Fig.~\ref{fig:magnetics}.  
The horizontal axis in these plots is the relative angle between the magnetic moments of Ir and Ni ions, $\phi_{\rm Ni-Ir}$.
In Fig.~\ref{fig:magnetics}(a) and (b), we plot the total energy per formula unit, and in Fig.~\ref{fig:magnetics}(c), 
we plot the angles $\phi_{\rm Ir}$ and $\phi_{\rm Ni}$ that the two magnetic moments make with the $z$ axis
(as defined in Fig.~\ref{fig:magneticstates}(e)). 
There is a clear trend in $\phi_{\rm Ir}$ and $\phi_{\rm Ni}$ as a function of $\phi_{\rm Ni-Ir}$.
The only ferromagnetic state ($\phi_{\rm Ni-Ir}=0$) is observed when $\phi_{\rm Ir}=\phi_{\rm Ni}=90^\circ$, consistent with the previous observation that 
we could stabilize FM only if the moments are in the $x$-$y$ plane (Fig.~\ref{fig:magneticstates}(b)). 
Similarly, ferrimagnetic order ($\phi_{\rm Ni-Ir}=180^\circ$) is observed
only for $\phi_{\rm Ir}=180^\circ$ and $\phi_{\rm Ni}=0^\circ$, i.e., only when the
moments are along $\mp \hat{z}$ (Fig.~\ref{fig:magneticstates}(a)). 
The intermediate data points in Fig.~\ref{fig:magnetics} correspond to intermediate states where the moments have their 
$z$ components ordered ferrimagnetically while the $x$ components are ordered ferromagnetically 
(Fig.~\ref{fig:magneticstates}(d)). 
Replacing the anisotropic interaction with an isotropic Heisenberg interaction and instead using the SIA 
to explain the Ising behaviour would result in intermediate 
states with antialigned moments tilted away from the high symmetry axes, such as those shown in Fig.~\ref{fig:magneticstates}(c). 
However, we never observed an intermediate state like this in Sr$_3$NiIrO$_6$, supporting the view that 
the interactions between the Ni and Ir ions are strongly anisotropic. 

Fitting the energy values to the anisotropic exchange model, we get 
$J_{\parallel}\!=\!19.0$\,meV/$\mu_B^2$ and $J_{\perp}\!=\!-8.4$\,meV/$\mu_B^2$. 
This simple model already fits the data quite well and gives the yellow data points in 
Fig.~\ref{fig:magnetics}(a). The discrepancy between the first-principles results and the model is due to both 
the numerical error and the presence of a nonzero SIA. Introducing SIA to the anisotropic exchange model 
makes a small additional improvement in the agreement between the DFT result and the model fit, 
as shown in Fig. \ref{fig:magnetics}(b). 

This is not the first study which uses an anisotropic exchange model for a compound with the 
K$_4$CdCl$_6$ structure. 
However, to the best of our knowledge, this is the first time that the exchange parameters $J_{\perp}$ and $J_{\parallel}$ are extracted
from first principles and microscopically justified for this compound. This is also the first prediction of opposite
signs for $J_{\perp}$ and $J_{\parallel}$.
Yin et al.\ have employed and microscopically justified a similar model 
to explain the magnetic anisotropy and magnon spectrum of Sr$_3$CuIrO$_6$.\cite{liu2012, yin2013} Also, 
both Toth et al.\cite{toth2016} and Lefrancois et al. used\cite{lefrancois2016} a similar model to explain their 
experimental observations of magnon spectra of Sr$_3$NiIrO$_6$. 
%
%
%
The connection between the magnetic anisotropy and exchange interactions were apparent 
in the results of Gordon et al.,\cite{gordon2016} who determined that the FM order is more stable when the spins are aligned 
in the $x$-$y$ plane, but their approach focused on the spin, not the pseudospin, of the Ir ion, and 
did not permit the construction of a simple magnetic Hamiltonian. 

We have intentionally refrained from introducing a SIA term into our model to emphasize that the 
physics of Sr$_3$NiIrO$_6$ \textit{can} be explained without it. 
A large SIA term is not physically justified in this compound: in a cubic environment 
Ni$^{2+}$ has two $e_g$ holes, and therefore no orbital angular momentum, and the $J_\textrm{eff}=1/2$ states of Ir  are SU(2) 
symmetric and therefore are not supposed to have any SIA. The trigonal crystal field necessarily 
breaks this simple picture, but there is no apparent reason why the trigonal field in this material should be strong enough to 
give rise to a record-breaking coercive field as well as a very large magnon gap. 
The anisotropic exchange interaction, on the other hand, leads to a magnetic anisotropy energy that is the same order of magnitude as 
the magnetic exchange itself, and can be used to explain the large observed coercive field.
%
%
The SIA is allowed by symmetry, and hence is definitely nonzero. However, our physical model, along with the first principles 
calculations, show that it is neither necessary to explain the experimental observations, nor the dominant source of 
anisotropy in Sr$_3$NiIrO$_6$. In other words, the \textit{minimal} model sufficient to explain all the experimental and 
theoretical observations does not require a SIA term, 
%
%
even though adding SIA improves the quality of the fit to the DFT energies (Fig. \ref{fig:magnetics}(b)). 
It is also possible to obtain an acceptable fit using a model with isotropic Heisenberg exchange and SIA; however, such a model does not
explain the theoretically observed magnetic configurations, and is not well motivated. 
(See the supplemental information for further discussion of different possible models and their fit to the DFT energies.) 

\section{Magnons} Magnons are commonly used to probe the nature of magnetic interactions. 
There are both inelastic neutron scattering (INS)
and resonant inelastic X-ray scattering (RIXS) experiments that probed 
the magnon spectrum of Sr$_3$NiIrO$_6$.\cite{toth2016,lefrancois2016} 
Even though each method is sensitive to only one of the two magnon branches in this compound,
together they present a coherent picture: 
One of the branches has a width of $\sim$10 meV, and is around $\sim$35 meV. The other branch, dominated by Ir, is at 
$\sim$90 meV, and is almost dispersionless. 
These observations of a large magnon splitting and gap, much larger than the bandwidth, 
have previously been
explained by a combination of anisotropic exchange, SIA, and 
Dzyaloshinskii-Moriya interactions.\cite{toth2016,lefrancois2016} Here we calculate the magnon 
spectrum of Sr$_3$NiIrO$_6$ using only the anisotropic exchange model with parameters from first principles to show that 
a model without SIA is sufficient to explain the large gap in the magnon 
spectrum.

\begin{figure}
\includegraphics[width=0.8\linewidth]{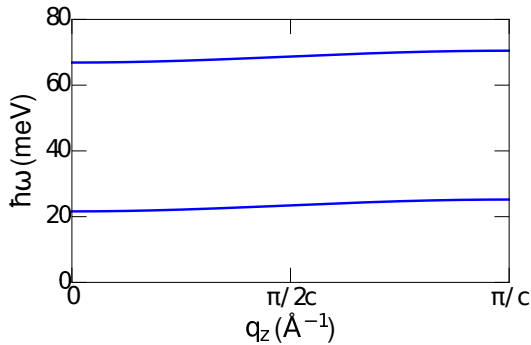}
\caption{(Color Online)
Magnon spectra of Sr$_3$NiIrO$_6$ for wavevectors along the [001] direction in the magnetically ordered phase, obtained 
from the anisotropic exchange model. }
\label{fig:magnons}
\end{figure}

We present the magnon spectra in Fig.~\ref{fig:magnons}.\cite{Note1} 
Our results correctly reproduce a large gap both between the two magnon branches, and between the lower 
branch and the zero-energy axis. The quantitative agreement is not perfect, but
his can be possibly fixed by fine-tuning the $U$ parameters. 

The sign of $J_\perp$ does not enter into the energy expression for the magnons, so  
the magnon spectra do not provide any evidence for the sign difference between $J_{\parallel}$ and $J_{\perp}$. 
%
%
However, there is a crucial effect of the radically anisotropic exchange: The characters (in-phase vs. out-of-phase) 
of the acoustic and optical modes at the zone center are flipped: the lower energy magnon branch has an oscillation 
pattern like in Fig. \ref{fig:magneticstates}d, and not like in Fig. \ref{fig:magneticstates}c.
As a result, the inelastic neutron scattering cross sections of these magnons are also flipped: 
While a precise calculation of the cross sections for an inelastic neutron scattering experiment is beyond the scope of this study,
in the supplemental information we provide a simple calculation that shows how the relative magnitudes of the magnon creation cross
sections of the two branches depend on the sign and not just the magnitude of $J_\perp$.
In other words, the relative intensities of the two branches carry information about the anisotropy in the 
exchange coupling, and experimental measurement of these intensities can provide the smoking gun evidence in support 
of the radically anisotropic exchange model. 
So far, the only neutron scattering study on this compound\cite{toth2016} could not observe the higher energy branch, and as a result 
there is no data to verify our prediction.

\section{Conclusions} 
The nearest neighbor magnetic interaction in the chain compound Sr$_3$NiIrO$_6$ 
is not an isotropic Heisenberg exchange, but is rather 
radically anisotropic: The different components of magnetic moments are coupled with opposite signs.  
This explains the observations of both the strong Ising-type behaviour and the large magnon gap without a
large, physically unjustified SIA, thus resolving the mystery of the large coercivity observed in this compound. 
The magnon frequencies do not contain a feature that can differentiate between the SIA and anisotropic exchange, 
but our model has a signature in the magnon creation cross sections. An experiment that can quantify these cross sections 
can conclusively differentiate between the radically anisotropic exchange and SIA. 

\begin{acknowledgements}
This work was supported by NSF DMREF Grant DMR-1629059.
\end{acknowledgements}

\end{document}